\documentclass[11pt]{article}

\usepackage[latin1]{inputenc}
\usepackage{graphicx}
\usepackage{color}
\usepackage{epsfig}
\usepackage{comment}
\usepackage{multirow}
\usepackage{amsmath}
\usepackage{amssymb}
\usepackage[hidelinks, bookmarksopen=true]{hyperref}
\usepackage{natbib}
\usepackage{booktabs}
\usepackage{longtable}
\usepackage{subfig}
\usepackage{enumitem}
\usepackage{footnote}
\usepackage{empheq}
\usepackage{amsthm}
\usepackage[mathscr]{euscript}

\newcommand{\xxi}{\mbox{\boldmath{$\xi$}}}
\newcommand{\mmu}{\mbox{\boldmath{$\mu$}}}
\newcommand{\ttheta}{\mbox{\boldmath{$\theta$}}}
\newcommand{\SSigma}{\mbox{\boldmath{$\Sigma$}}}
\newcommand{\1}{\bf 1}
\newcommand{\0}{\bf 0}
\newcommand{\oo}{\bf o}
\newcommand{\rr}{\bf r}
\newcommand{\uu}{\bf u}
\newcommand{\vv}{\bf v}

\newcommand{\D}{\bf D}
\newcommand{\I}{\bf I}
\newcommand{\Q}{\bf Q}
\newcommand{\W}{\bf W}

\begin{document}


\title{\textbf{Scalable Bayesian modeling for smoothing disease risks in large spatial data sets}}

\author{Orozco-Acosta, E$^{1,2}$, Adin, A$^{1,2}$, and Ugarte, M.D.$^{1,2}$\\
\small {\textit{$^1$ Department of Statistics, Computer Sciences and Mathematics, Public University of Navarre, Spain.}} \\
\small {\textit{$^2$ Institute for Advanced Materials and Mathematics, InaMat$^2$, Public University of Navarre, Spain.}}\\
\small { $*$Correspondence to María Dolores Ugarte, Departamento de Estad\'istica, Inform\'atica y Matem\'aticas, } \\
\small { Universidad P\'ublica de Navarra, Campus de Arrosadia, 31006 Pamplona, Spain.} \\
\small {\textbf{E-mail}: lola@unavarra.es }}
\date{}

\makeatletter
\pdfbookmark[0]{\@title}{title}
\makeatother

\maketitle

\begin{abstract}

Several methods have been proposed in the spatial statistics literature for the analysis of big data sets in continuous domains. However, new methods for analyzing high-dimensional areal data are still scarce. Here, we propose a scalable Bayesian modeling approach for smoothing mortality (or incidence) risks in high-dimensional data, that is, when the number of small areas is very large. The method is implemented in the R add-on package\verb" bigDM". Model fitting and inference is based on the idea of ``divide and conquer" and use integrated nested Laplace approximations and numerical integration. We analyze the proposal's empirical performance in a comprehensive simulation study that consider two model-free settings. Finally, the methodology is applied to analyze male colorectal cancer mortality in Spanish municipalities showing its benefits with regard to the standard approach in terms of goodness of fit and computational time.

\end{abstract}

Keywords: High-dimensional data; INLA; Hierarchical models; Mixture models; Spatial epidemiology
\bigskip

\section{Introduction}
\label{sec:Introduction}

Statistical models are an essential tool for the analysis of the geographical or spatial distribution of environmental and epidemiological data in small areas. Nowadays, one of the biggest challenge in the field of spatial statistics is the development of new computationally efficient methods that are able to obtain reliable estimates of the underlying geographical pattern for large data sets. Several modern methods have been proposed for the analysis of massive geostatistical (point-referenced) data, where traditional estimation of Gaussian processes (GPs) becomes computationally prohibitive. Some of these approaches includes low-rank approximations to GPs such as fixed-rank kriging \cite{cressie2008fixed}, predictive processes \cite{banerjee2008gaussian}, stochastic partial differential equations \cite{lindgren2011explicit}, lattice kriging \cite{nychka2015multiresolution}, multi-resolution approximations \cite{katzfuss2017multi}, and Vecchia approximations \cite{datta2016hierarchical,katzfuss2019general} among others, plus several parallel computation algorithmic approaches as for example \cite{gramacy2015local,paciorek2015parallelizing,guhaniyogi2018meta,katzfuss2017parallel}. However, there is not much research about the scalability of statistical models for areal (lattice) data.

Disease mapping is the field of spatial epidemiology that studies the link between geographic locations and the occurrence of diseases, focusing on the estimation of the spatial and/or spatio-temporal distribution of disease incidence or mortality patterns \cite{lawson2016handbook,martinez2019disease}. In these studies the region of interest is divided in non-overlapping irregular areal units (administrative divisions such as states or local health areas), where epidemiological data are presented as aggregated disease counts for each geographical unit. The great variability inherent to classical risk estimation measures, such as standardized mortality/incidence ratios or crude rates, makes necessary the use of statistical models to smooth spatial risk surfaces. Bayesian hierarchical models are typically used for such objective, where spatially structured random effects are included at the second level of the hierarchy. Most of the research in spatial disease mapping is based on the conditional autoregressive (CAR) prior distribution \cite{besag1974spatial}, where the spatial correlation between random effects is determined by the neighbouring structure (represented as an undirected graph) of the areal units. Despite the enormous expansion of modern computers and the development of new software and estimation techniques to make fully Bayesian inference, dealing with high-dimensional spatial random effects is still computationally challenging.

As far as we know, there are very few works in the disease mapping literature that proposed computationally efficient methods to analyze large-scale spatial data. Hughes and Haran (2013) \cite{hughes2013dimension} propose a parameterization of the areal spatial generalized linear mixed model that alleviates spatial confounding when including covariates in the model (see for example \cite{reich2006effects} and \cite{hodges2010adding}) while speading computation by greatly reducing the dimension of the spatial random effect. To achieve this dimension reduction, they suggest to reparameterize the model selecting a fixed number of eigenvectors of the Moran operator (those corresponding to the largest eigenvalues to include patterns of positive spatial dependence, i.e., attraction, or those corresponding to the smallest eigenvalues to include patterns of negative spatial dependence, i.e., repulsion). The model is implemented in the \texttt{R} package \texttt{ngspatial} \cite{ngspatial}.
Very recently, Datta et al. (2019) \cite{datta2019} propose a new way of constructing precision matrices for count data models using a directed acyclic graph representation derived from the original spatial neighbourhood structure of the areal units. Instead of modeling the precision matrix of the spatial random effect directly, they propose to model its (sparse) Cholesky factor using autoregressive covariance models on a sequence of local trees created from this directed acyclic graph. Although the proposed model is order-dependent, as stated by the own authors,  the joint density of the spatial random effect will be scalable for large datasets.

In this paper, we propose a scalable Bayesian modeling approach for smoothing mortality (or incidence) risks for high-dimensional spatial disease mapping data, that is, when the number of small areas is very large. Our method is based on the well-known ``divide and conquer'' approach. Instead of considering a global spatial random effect whose correlation structure is based on the whole neighbourhood graph of the areal units, the spatial domain is divided into $D$ subregions so that local spatial models can be fitted simultaneously (in parallel). Two different models are proposed based on the partition of the geographical units. The first model assumes that the spatial domain is divided into $D$ disjoint subregions, according to administrative subdivisions for example. Then, independent spatial models are fitted to each data subset based on the neighbourhood structure of the corresponding subgraphs. Once computations are finished, the area-specific relative risks are merged to obtain a single spatial risk surface. Clearly, assuming independence between areas corresponding to different subregions of the partition of the spatial domain could lead to border effects in risk estimates. To avoid this undesirable issue, we also propose a second modeling approach where $k$-order neighbours are added to each subregion of the spatial domain. In consequence, the main spatial domain is divided into overlapping partitions.  This means that  some areal units will have several risk estimates. To obtain a unique posterior distribution for these risks, we propose to compute a mixture distribution of the estimated posterior probability density functions of the risks. In addition, approximate values for some model selection criteria are derived to perform Bayesian model comparison.

A simulation study is conducted to compare our scalable model's proposals against the global model using the almost 8000 municipalities of continental Spain. This study reveals a competitive performance of the new model proposals in terms of goodness of fit and computational time, that is reduced substantially. We observe that as we increase the neighbourhood ordering ($k$ parameter) in our second modeling approach, the results are more similar to the global model, but this comes with a loss of computational efficiency. The new methodology will be used to analyse male colorectal cancer mortality in Spanish municipalities.

The rest of the paper is organized as follows. In Section \ref{sec:SpatialModels} we briefly review some spatial models in disease mapping and we give some details about Bayesian inferential techniques to fit these models. Section \ref{sec:Models} introduces the scalable model proposals for high-dimensional areal count data described in this work. In Section \ref{sec:SimulationStudy} a simulation study is conducted to compare the performance of our modeling approach with the usual spatial model for areal count data. Male colorectal cancer mortality data in Spanish municipalities is analyzed in Section \ref{sec:DataAnalysis}. The paper concludes with a discussion and some conclusions.
The methods and algorithms proposed here are implemented in the \texttt{R} package \texttt{bigDM} available at \url{https:
https://github.com/spatialstatisticsupna/bigDM}, which contains a vignette to replicate the data analysis described in this paper using a simulated colorectal cancer mortality data (modified in order to preserve the confidentiality of the original data).

\section{Spatial models for disease mapping}
\label{sec:SpatialModels}

Let us assume that the spatial domain of interest is divided into $n$ contiguous small areas labeled as $i=1,\ldots,n$. For a given area $i$, $O_i$ will denote the observed number of disease cases and $N_i$ the population at risk. The simplest mortality/incidence indicator is the \textit{crude rate}, which is usually defined as the number of cases per 100,000 people, that is, $CR_i=\frac{O_i}{N_i} \times 100,000$. When the aim of the study is to detect which areas exhibit elevated or lowered risk, the number of expected cases  in each small area are usually computed. For example, if the population is divided into age-groups, the indirect standardization method is commonly  used  to calculate the expected number of cases as $E_i=\sum_{j=1}^J N_{ij} \frac{O_j}{N_j}$ for $i=1,\ldots,n$, where $O_j=\sum_{i=1}^n O_{ij}$ and $N_j=\sum_{i=1}^n N_{ij}$ are the number of cases and the population at risk in the $j$th age-group, respectively. Note that $E_i$ represents the number of cases we expect to observe in the $i$th area if it behaves as the whole study region. Using these quantities, the \textit{standardized mortality/incidence ratio} (SMR or SIR) is defined as the ratio of observed and expected cases for the corresponding areal unit. Although its interpretation is very simple (areas with values higher than 1 will stand for an excess of risk, while values lower than 1 mean a lower risk for the population in that unit), these measures are extremely variable when analyzing rare diseases or low-populated areas, as it is the case of high-dimensional data. To cope with this situation, it is necessary to use statistical models that stabilize the risks (rates) borrowing information from neighbouring regions.

Generalized linear mixed models (GLMM) are typically used for the analysis of count data within a hierarchical Bayesian framework. Conditional to the relative risk $r_i$, the number of observed cases in the $i$th area is assumed to be Poisson distributed with mean $\mu_{i}=E_{i}r_{i}$. That is, %
\begin{eqnarray*}
\label{eq:Model_Poisson}
\begin{array}{rcl}
O_{i}|r_{i} & \sim & Poisson(\mu_{i}=E_{i}r_{i}), \ i=1,\ldots,n\\
\log \mu_{i} & = & \log E_{i}+\log r_{i},
\end{array}
\end{eqnarray*}
where $\log E_{i}$ is an offset. Depending on the specification of the log-risks different models are defined. Here we assume that
\begin{equation}
\label{eq:Model1}
\log r_{i}=\alpha+\xi_{i},
\end{equation}
where $\alpha$ is an intercept representing the overall log-risk and $\xi_i$ is a spatial random effect. Commonly, a conditional autoregressive (CAR) prior distribution is assumed for the random effect $\xxi=(\xi_1,\ldots,\xi_n)^{'}$, which is a type of Gaussian Markov random field (GMRF) \cite{rue2005gaussian}. A GMRF, with respect to a given graph, is defined on a vector $\xxi$ by assuming a multivariate Normal distribution $\xxi \sim N(\mmu,\SSigma)$, where ${\SSigma}^{-1}=\Q$ is a $n \times n$ sparse precision matrix corresponding to the undirected graph of the regions under study.
In what follows, we briefly review some of the most commonly used CAR priors for spatial random effects. Let ${\W}=(w_{ij})$ be a binary $n \times n$ adjacency matrix, whose $ij$th element is equal to one if areas $j$ and $k$ are defined as neighbours, usually if they share a common border (denoted as $i \sim j$), and it is zero otherwise. The joint distribution of the \textit{intrinsic CAR} prior (iCAR) \cite{besag1991} is defined as
\begin{equation*}
\label{eq:iCAR}
\xxi \sim N({\0},{\Q}^{-}_{\xi}), \quad \mbox{with} \quad {\Q}_{\xi}=\tau_{\xi}(\D_{W}-\W)
\end{equation*}
where ${\D}_{W} = diag(w_{1+},\ldots,w_{n+})$ and $w_{i+}=\sum_j w_{ij}$ is the $i$th row sum of $\W$, and $\tau_{\xi}=1/\sigma^2_{\xi}$ is the precision parameter. As ${\Q}_{\xi}{\1}_n={\0}$, where ${\1}_n$ is a vector of ones of length $n$ (i.e., ${\1}_n$ is the eigenvector associated to the null eigenvalue of ${\Q}_{\xi}$), the precision matrix of the iCAR distribution is singular and therefore, the joint distribution of $\xxi$ is improper. If the spatial graph is fully connected (matrix ${\Q}_{\xi}$ has rank-deficiency equal to 1), a sum-to-zero constraint $\sum_{i=1}^n \xi_i = 0$ is usually imposed to solve the identifiability issue between the spatial random effect and the intercept in Model \eqref{eq:Model1}.

The iCAR prior distribution only accounts for spatial correlation structures, and hence, it is not appropriate if the data variability  is not only spatially structured but unstructured heterogeneity is also present. A \textit{convolution} prior was also proposed by \cite{besag1991} to deal with this situation (usually named as BYM prior) that combines the iCAR prior and an additional set of unstructured random effects. The model is given by
\begin{equation*}
\label{eq:BYM}
\xxi = \uu + \vv, \quad \mbox{with} \quad \begin{array}{l} {\uu} \sim N({\0},[\tau_{u}({\D}_{W}-{\W})]^{-}), \\ {\vv} \sim N({\0},\tau_{v}^{-1}{\I}_n). \end{array}
\end{equation*}
where ${\I}_n$ is the $n \times n$ identity matrix. The precision parameters of the spatially structured random effect ($\tau_{u}$) and the unstructured random effect ($\tau_{v}$) are not identifiable from the data \cite{macnab2011gaussian}, just the sum $\xi_i=u_i + v_i$ is identifiable. Hence, similar to the iCAR prior distribution, the sum-to-zero constraint $\sum_{i=1}^n (u_i+v_i) = 0$ must be imposed to solve identifiability problems with the intercept.

Leroux et al. \cite{leroux1999estimation} propose an alternative proper CAR prior (hereafter named as LCAR prior) to model both spatially structured and unstructured variation in a single set of random effects. It is given by
\begin{equation*}
\label{eq:LCAR}
\xxi \sim N({\0},{\Q}^{-1}_{\xi}), \quad \mbox{with} \quad {\Q}_{\xi}=\tau_{\xi}[\lambda_{\xi}({\D}_W-{\W})+(1-\lambda_{\xi}){\I}_n]
\end{equation*}
where $\tau_{\xi}$ is the precision parameter and $\lambda_{\xi} \in [0,1)$ is a spatial smoothing parameter. Even the precision matrix ${\Q}_{\xi}$ is of full rank whenever $0 \leq \lambda_{\xi} < 1$, a confounding problem still remains and consequently, a sum-to-zero constraint $\sum_{i=1}^n \xi_i = 0$ has to be considered (see \cite{goicoa2018spatio}).

Other conditional autoregressive priors have been also given in the literature, like the \textit{proper CAR} prior distribution described in \cite{cressie1993statistics}, or the reparameterization of the BYM model given by Dean et al. \cite{dean2001detecting}.

\subsection{Model fitting and inference}
\label{sec:INLA}

The fully Bayesian approach is probably the most-used technique for model fitting and inference in spatial disease mapping. Under this framework, the entire posterior probability distribution for the parameters of interest are obtained. Traditionally, Markov chain Monte Carlo (MCMC) techniques have been used for model inference from a fully Bayes perspective, mainly due to the development and accessibility of the well-known \texttt{WinBUGS} \cite{spiegelhalter2003winbugs} software. 
During the last years, other softwares based on MCMC methods are being popularized such as JAGS \cite{plummer2003jags} or STAN \cite{stan2018stan}, as well as other new statistical systems as NIMBLE \cite{nimble}.
An alternative to MCMC simulation methods for Bayesian inference was proposed by \cite{rue2009approximate}. The method known as INLA is based on integrated nested Laplace approximations and numerical integration. The main goal of the INLA strategy is to approximate the marginal posterior distribution of a GMRF using numerical methods for sparse matrices to speed up computations in comparison with MCMC methods. This technique can be used easily in the free software \texttt{R} through the \texttt{R-INLA} package (\url{http://www.r-inla.org/}). The use of INLA for Bayesian inference has turned out to be very popular in applied statistics in general, and in the field of spatial statistics in particular. A review of the INLA method and references to some of its more recent applications can be found in \cite{rue2017bayesian}.

Despite the computational efficiency of INLA for Bayesian inference when fitting spatial and spatio-temporal disease mapping models for areal data, its use has not been studied in detail when the number of areas increases considerably. New parallelization strategies have been recently implemented in INLA through the integration of a special version of the PARDISO (\url{www.pardiso-project.org}) library \cite{van2019new}. However, the computational resources needed for analyzing massive spatial data could be enormous, something that is not within the reach of all researchers in statistics, epidemiologists or public health professionals.
Thus, the main objective of this paper is to provide an alternative scalable method to perform high-dimensional spatial analysis for count data with INLA. Although the methodology described in the next section is focused on the INLA estimation strategy, it can be also adapted to other Bayesian estimation techniques.

\section{Scalable Bayesian model proposal}
\label{sec:Models}

In this section, we propose a scalable Bayesian modeling approach for smoothing mortality (or incidence) risks for high-dimensional spatial disease mapping data. Our proposal is based on applying the ``divide and conquer'' approach to the spatial model described in Equation \eqref{eq:Model1}, which will be named as the \textit{Global model}. The key idea is to divide the spatial domain into $D$ subregions so that local spatial models can be simultaneously fitted in parallel reducing the computational time substantially. The LCAR prior distribution has been considered for the spatial random effect ${\xxi}$, but any other CAR distribution as those described in Section \ref{sec:SpatialModels} could be used instead in the methodology described below.

\subsection{Disjoint models}
\label{sec:Models_Disjoint}

Let consider a partition of the spatial domain $\mathscr{D}$ into $D$ subregions, that is $\mathscr{D} = \bigcup_{d=1}^D \mathscr{D}_d$ where $\mathscr{D}_i \cap \mathscr{D}_j = \emptyset$ for all $i \neq j$. In our disease mapping context, this means that each geographical unit belongs to a single subregion. A natural choice for this partition could be the administrative subdivisions of the area of interest (such as for example, provinces or states).

Let ${\bf O}_d = \{O_{i} | \mbox{ area } i \in \mathscr{D}_d \}$ and ${\bf E}_d = \{E_{i} | \mbox{ area } i \in \mathscr{D}_d \}$ represent the observed and expected number of disease cases in each subregion, respectively. It is important to remark that the expected values are computed using all the data. Then, for $d=1,\ldots,D$ the log-risks of the  \textit{Disjoint models} are expressed in matrix form as
\begin{eqnarray}
\label{eq:Model2}
\begin{array}{rcl}
\log {\rr}_d & = & {\alpha}_d + {\xxi}_d,\\[1.ex]
{\xxi}_d & \sim & N\left({\0},[\tau_{\xi_d}(\lambda_{\xi_d}({\D}_{W_d}-{\W}_d)+(1-\lambda_{\xi_d}){\I}_{n_d})]^{-1}\right)
\end{array}
\end{eqnarray}
where $\alpha_d$ is an intercept, ${\xxi}_d = (\xi_1^{d},\ldots,\xi_{n_d}^{d})^{'}$ is the vector of spatial random effects within each subregion with a LCAR prior distribution, ${\W}_d$ is the neighbourhood subgraph of the areas belonging to $\mathscr{D}_d$, and ${\I}_{n_d}$ is the identity matrix of dimension $n_d$, with $\sum_{d=1}^D n_d = n$. Note that this model can be also written as
\begin{equation*}
\label{eq:Model2_ref}
\begin{pmatrix}
\log {\rr_1} \\ \vdots \\ \log {\rr_{D}}
\end{pmatrix}
=
\begin{pmatrix}
{\1}_{n_1} & & \\ & \ddots & \\ & & {\1}_{n_D}
\end{pmatrix}
\begin{pmatrix}
{\alpha}_1 \\ \vdots \\ {\alpha}_D
\end{pmatrix}
+
\begin{pmatrix}
{\I}_{n_1} & & \\ & \ddots & \\ & & {\I}_{n_D}
\end{pmatrix}
\begin{pmatrix}
{\xxi}_1 \\ \vdots \\ {\xxi}_D
\end{pmatrix}
\end{equation*}
where ${\1}_{n_d}$ are column vectors of ones of length $n_d$, and the precision matrix of the multivariate Normal random effect vector ${\xxi} = ({\xxi}_1,\ldots,{\xxi}_D)^{'}$ is a block-diagonal matrix of dimension $n \times n$ with blocks corresponding to the precision matrix of the LCAR prior within each subgraph. However, under the formulation of Equation \eqref{eq:Model2}, $D$ independent spatial models can be simultaneously fitted giving rise to a clear computational gain.

Since we have defined a partition of the spatial domain $\mathscr{D}$, the log-risk surface $\log {\rr} = (\log {\rr_1},\ldots,\log {\rr_{D}})^{'}$ is just the union of the posterior estimates of each submodel. However, note that $D$ specific intercepts are estimated in Model \eqref{eq:Model2}. To obtain a single estimate of the overall log-risk $\alpha$ as in Model \eqref{eq:Model1}, we propose to extract samples from the joint posterior distribution of the linear predictors $\log {\rr_d}$ (for $d=1,\ldots,D$) using the \texttt{inla.posterior.sample()} function of \texttt{R-INLA}. This function allows to generate samples from the approximate joint posterior marginal of a previously fitted \textit{inla} object, if the argument \texttt{control.compute = list(config = TRUE)} is provided when calling the \texttt{inla()} function (see for example, \cite{gomez2020bayesian} and \cite{martino2019integrated}). After joining the $S$ samples from each submodel, we define
\begin{equation*}
\alpha^s = \frac{1}{n} \sum_{i=1}^n \log r_i, \quad \mbox{for } s=1,\ldots,S
\end{equation*}
and then compute the kernel density estimate of $\alpha$ \cite{sheather1991reliable}. 

\subsection{$k$-order neighbourhood model}
\label{sec:Models_K-order}
Assuming independence between areas belonging to different subregions could be very restrictive and may lead to border effects in the disease risk estimates. To avoid this undesirable issue, we also propose a second modeling approach where $k$-order neighbours are added to each subregion of the spatial domain. Notice that doing this, the main spatial domain $\mathscr{D}$ is now divided into overlapping set of regions, that is, $\mathscr{D} = \bigcup_{d=1}^D \mathscr{D}_d$ but $\mathscr{D}_i \cap \mathscr{D}_j \neq \emptyset$ for neighbouring subregions. In consequence, for some areal units multiple relative risk estimates will be obtained. As in the disjoint model of Equation \eqref{eq:Model2}, $D$ submodels will be simultaneously fitted using \texttt{R-INLA}. However, the final risk surface ${\rr} = (r_1,\ldots,r_n)^{'}$ is no longer the union of the posterior estimates obtained for each submodel, since $\sum_{d=1}^D n_d > n$. 

To obtain a unique posterior distribution of $r_i$ for each areal unit $i$, we propose to compute a mixture distribution (see, e.g., \cite{lindsay1995mixture,fruhwirth2006finite})  using  the estimated posterior probability density function of these risks. Let us assume that area $i$ lies within $m(i)$ subregions of the spatial domain $\mathscr{D}$. That is, we have $m(i)$   estimates of the $i$th area risk. If we denote $f_1(x), \ldots, f_{m(i)}(x)$ to the posterior estimates of the probability density functions, the mixture distribution of $r_i$ can be written as the weighted sum of the corresponding densities
\begin{equation*}
f(x)=\sum_{j=1}^{m(i)} w_jf_j(x),
\end{equation*}
where $w_j \geq 0$ and $\sum_{j=1}^{m(i)} w_j= 1$. The approximate posterior density functions $f_j(x)$ are obtained from the corresponding submodels using the \texttt{inla.dmarginal()} function. We propose to use the \textit{conditional predictive ordinate} (CPO), a diagnostic measure to detect discrepant observations from a given model \cite{pettit1990conditional}, to compute the weights of the mixture distribution dividing each CPO value by the sum for the $m(i)$ different estimates. Note that giving the set of observations ${\oo}=(o_1,\ldots,o_n)^{'}$, $\mbox{CPO}_i = Pr(O_i = o_i | {\oo}_{-i})$ values denotes the cross-validated predictive probability mass at the observed count $o_{i}$. As described in \cite{rue2009approximate}, the CPO quantities are computed in \texttt{R-INLA} without re-running the model by including into the \texttt{inla()} function the argument \texttt{control.compute=list(cpo=TRUE)}.
%

\subsection{Model selection criteria}
\label{sec:Models_DIC}

In this section we discuss some Bayesian model selection criteria  and show how to compute them when fitting disjoint and $k$-order neighbourhood models.
Given the data ${\oo}$ with likelihood function $p({\oo}|{\ttheta})$ where ${\ttheta}$ are the unknown parameters of the model, the \textit{Bayesian deviance} is defined as
\begin{equation*}
\label{eq:Deviance}
D({\ttheta}) = -2 \log(p({\oo}|{\ttheta})) + 2 \log p({\oo})
\end{equation*}
where $2 \log p({\oo})$ denotes the deviance of the saturated model (a constant that does not depend on the model parameters). Note that under our model formulation, that is $O_i | r_i \sim Poisson(\mu_i=E_i r_i)$, the log-likelihood function is expressed as
\begin{equation*}
\label{eq:Poisson_likelihood}
\log(p({\oo}|{\ttheta})) = \log\left(\prod_{i=1}^{n}\frac{e^{-\mu_{i}}\mu_{i}^{o_{i}}}{o_{i}!}\right) =
\sum_{i=1}^n \log \left(\frac{e^{-\mu_{i}}\mu_{i}^{o_{i}}}{o_{i}!}\right).
\end{equation*}
Generally, the posterior mean deviance $\overline{D({\ttheta})}$ is considered as a measure of goodness of fit due to its robustness. However more complex models will fit the data  better, and consequently lower values of the mean deviance will be obtained. To avoid selecting models that overfit the data, several criteria that also take into account the model complexity have been proposed in the literature. Probably, the \textit{deviance information criterion} (DIC) \cite{spiegelhalter2002bayesian} and \textit{Watanabe-Akaike information criterion} (WAIC) \cite{watanabe2010asymptotic}, are two of the most well-known criteria to compare models in a fully Bayesian setting.

The DIC is computed as the sum of the posterior mean of the deviance and the number of effective parameters (a measure of model complexity)
\begin{equation*}
\label{eq:DIC1}
\mbox{DIC} = \overline{D({\ttheta})} + p_D,
\end{equation*}
where the quantity $p_D$ is defined as the posterior mean of the deviance minus the deviance computed at the posterior mean of the parameters of interest, thus,
\begin{equation*}
\label{eq:DIC2}
\mbox{DIC} = \overline{D({\ttheta})} + (\overline{D({\ttheta})}-D(\bar{\ttheta})) = 2\overline{D({\ttheta})}-D(\bar{\ttheta}).
\end{equation*}
Analogously to the Akaike information criterion (AIC), models with smaller DIC values provide better trade-off between model fit and complexity. To compute the DIC values in \texttt{R-INLA} for the \textit{Global model} described in Equation \eqref{eq:Model1}, the option \texttt{control.compute = list(dic = TRUE)} inside the \texttt{inla()} function is used. However, in order to compare this model with the scalable model proposals described in Sections \ref{sec:Models_Disjoint} and \ref{sec:Models_K-order}, approximate DIC values are computed for the latter models by drawing samples from the posterior marginal distributions of the relative risks using the \texttt{inla.rmarginal()} function. If a total of $S$ samples are drawn from each $r_i$, and denoting as ${\ttheta}^{s}$ to the posterior simulations of $\mu_i=E_ir_i$ for $s=1,\ldots,S$, we can compute approximate values of the mean deviance $\overline{D({\ttheta})}$ and the deviance of the mean $D(\bar{\ttheta})$ as
\begin{eqnarray*}
\label{eq:DIC_approx}
\begin{array}{l}
\overline{D({\ttheta})} \approx \frac{1}{S} \sum\limits_{s=1}^S -2 \log(p({\oo}|{\ttheta}^s)), \\[2ex]
D(\bar{\ttheta}) \approx -2 \log(p({\oo}|\bar{\ttheta})), \quad \mbox{with } \bar{\ttheta}=\frac{1}{S}\sum\limits_{s=1}^S {\ttheta}^s. \\
\end{array}
\end{eqnarray*}

Similarly, to compute the WAIC values in \texttt{R-INLA}, the option \texttt{control.compute=list(waic=TRUE)} must be used when fitting the \textit{Global model}. Following \cite{gelman2014understanding}, approximate WAIC values have been also computed for the \textit{Disjoint model} and the \textit{$k$-order neighbourhood model} as
\begin{equation*}
\label{eq:WAIC_approx}
\begin{array}{rcl}
\mbox{WAIC} & \approx & -2 \sum\limits_{i=1}^{n}\log \left(\frac{1}{S}\sum\limits_{s=1}^{S} p(o_{i}|{\ttheta}^s)\right) \\[2ex]
& & + 2 \sum\limits_{i=1}^{n} \mbox{Var}\left[ \log(p({o_i}|{\ttheta}^s)) \right].
\end{array}
\end{equation*}

\section{Simulation study}
\label{sec:SimulationStudy}
\begin{figure}
\vspace{-0.5cm}
\includegraphics[width=0.5\textwidth]{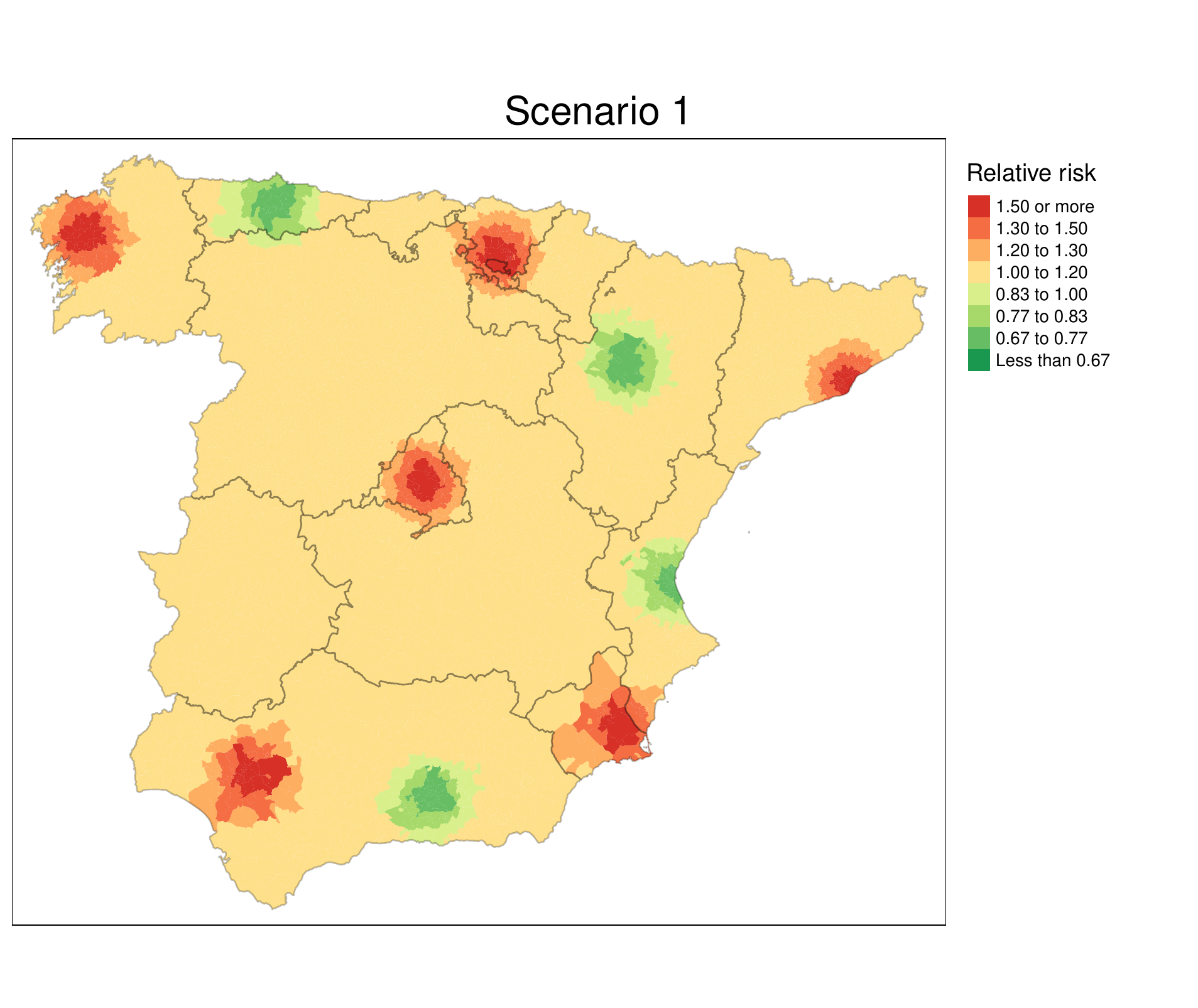}\hfill
\includegraphics[width=0.5\textwidth]{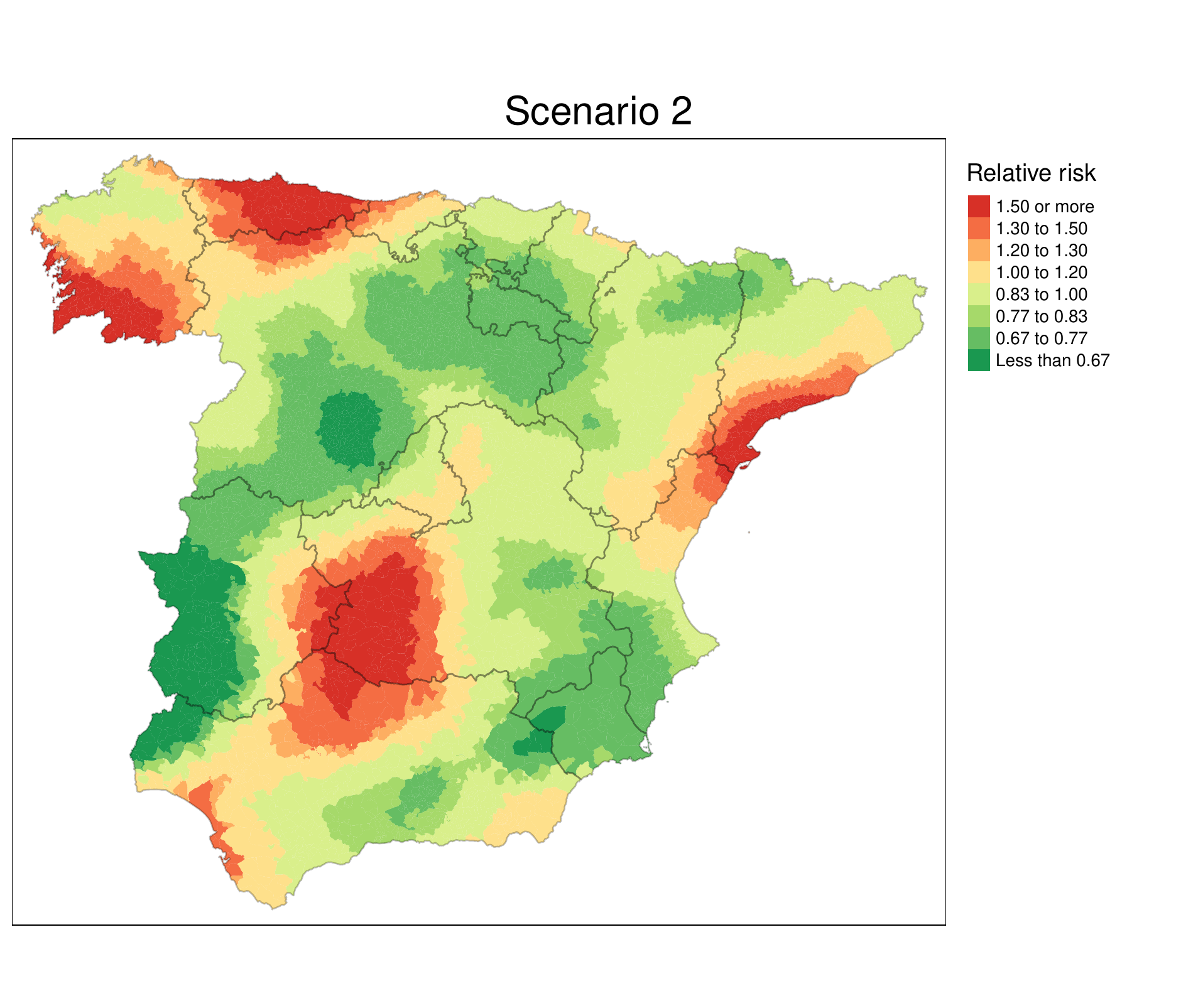}
\vspace{-0.5cm}
\caption{True risk surfaces for the simulation study of Scenario 1 (left) and Scenario 2 (right)}
\label{fig:True_risks}
\end{figure}

In this section, a simulation study is conducted to compare the scalable model proposals, i.e., the \textit{Disjoint model} described in Equation \eqref{eq:Model2} and the \textit{k-order neighbourhood model} described in Section \ref{sec:Models_K-order}, against the common spatial LCAR model described in Equation \eqref{eq:Model1}, denoted as \textit{Global model}. We base our study on the $n=7,907$ municipalities of continental Spain. To imitate the real case study that is analyzed in the next section, the $D=15$ Autonomous Regions of Spain are used as a partition of the spatial domain (see \autoref{fig:True_risks}).

To fit the models, improper uniform prior distributions are given to all the standard deviations (square root inverse of precision parameters), and a Uniform $(0,1)$ distribution is considered for the spatial smoothing parameters of the LCAR prior. Finally, a vague zero mean normal distribution with a precision close to zero (0.001) is given to the intercept ($\alpha$). All the calculations are made on a twin superserver with four processors, Inter Xeon 6C and 96GB RAM, using the full Laplace approximation strategy in \texttt{R-INLA} (stable) version INLA\_19.09.03 of R-3.6.2.

We consider two different scenarios to compare the performance of the models. In the first scenario, a model-free true risk surface is defined by randomly assigning high and low risk values to the areas surrounding some selected major cities of Spain. Considering these cities as the area' centroids, the relative risks are gradually increased/decreased at different distances to get  smooth surface. Specifically, relative risks of 1.5, 1.3 and 1.2 are assigned to the municipalities that are at less than 15km, 30km and 45km respectively from the centroids selected as high-risk areas. The same criterion has been used to assign reciprocal risks of 0.67, 0.77 and 0.83 to the municipalities surrounding a low-risk centroid. In the second scenario, a smooth risk surface is generated by sampling from a two-dimensional isotropic P-spline model with 40 equally spaced knots for longitude and latitude. The true risk surfaces for these scenarios are displayed in \autoref{fig:True_risks}.

In both scenarios, counts for each municipality are generated using a Poisson distribution with mean $\mu_i=E_ir_i$, where the number of expected cases $E_i$ are fixed at values equal to 1, 5, 10, and 50. A total of 100 simulations have been generated for each of the eight sub-scenarios.

\subsection{Results}
\label{sec:SimulationStudy_Results}
We evaluate the models' performance in terms of relative risk estimates by computing the mean absolute relative bias (MARB) and mean relative root mean square error (MRRMSE), defined as
\begin{equation*}
\begin{array}{rcl}
\mbox{MARB} & = & \frac{1}{n}\sum\limits_{i=1}^n \frac{1}{100} \left| \sum\limits_{l=1}^{100} \frac{\hat{r}_i^l-r_i}{r_i} \right|,\\[3.ex]
\mbox{MRRMSE} & = & \frac{1}{n}\sum\limits_{i=1}^n \sqrt{\frac{1}{100} \sum\limits_{l=1}^{100} \left(\frac{\hat{r}_i^l-r_i}{r_i} \right)^2},
\end{array}
\end{equation*}
where $r_i$ are the true generated risk, and $\hat{r}_i^l$  are the posterior median estimate of the relative risk for areal unit $i$ in the $l$-th simulation. In addition, coverage probabilities and 95\% credible intervals's lengths have been computed.

The average values for the 100 simulated datasets in each of the sub-scenarios are computed in \autoref{tab:SimulationStudy_Results}. The \textit{3rd order neighbourhood model} was also considered (not shown in the table), but results did not improve those obtained with lower neighbourhood orders. Regarding computational times (in seconds), those corresponding to models simultaneously fitted in multiple machines (T1) or in a single machine (T2) are included. The maps with average values of relative risk estimates for each sub-scenario are shown in the online supplementary material.

When the number of expected cases is very low, as in sub-scenarios with E=1, both model selection criteria and risk estimation accuracy measures, point out the \textit{Global model} as the best candidate. However, small differences are observed between this model and the \textit{1st order neighbourhood model}. As the number of expected cases increases, lower DIC/WAIC and better values of MARB and MRRMSE are observed for our scalable model proposals in Scenario 1. The \textit{1st order neighbourhood model} shows better values in terms of model selection criteria for sub-scenarios E=5, 10, and 50. Since in this scenario most of the high/low risk ``clusters'' are located inside the frontiers of the autonomous regions (see \autoref{fig:True_risks}), the performance of the \textit{Disjoint model} is also pretty good in terms of \sloppy MRRMSE.
However, Scenario 2 shows a much more extended risk surface across the whole spatial domain. That is the reason why, the \textit{Disjoint model} performs worse than the \textit{k-order neighbourhood models}, which are able to better recover the true risk surface. In sub-scenarios E=1, 5 and 10 the models with $k=2$ shows slightly smaller values of DIC and WAIC than models with first order neighbourhoods.

In general, we think that the new  modeling proposals are a very competitive alternative to the \textit{Global model} with a significant gain in computational time without a remarkable difference in terms of bias and variability.  Empirical coverages and credible interval lengths are in general very similar.

\begin{table}
\caption{Average values of deviance information criterion (DIC), Watanabe-Akaike information criterion (WAIC), mean absolute relative bias (MARB), mean relative root mean square error (MRRMSE), empirical coverage, length of the 95\% credible interval for the risks, and computational times (T1: approximate value of CPU time if all submodels are simultaneously fitted in multiple machines, T2: CPU time if all submodels are fitted in a single machine) in seconds.}
\label{tab:SimulationStudy_Results}
\resizebox{\textwidth}{!}{
\begin{tabular}{llrrrrrrrrrr}
\hline\noalign{\smallskip}
& & \multicolumn{2}{c}{Model selection criteria} & & \multicolumn{4}{c}{Risk estimation evaluation} & & \multicolumn{2}{c}{Time}\\
\cline{3-4} \cline{6-9} \cline{11-12}\\
& Model & DIC & WAIC & & MARB & MRRMSE & Cov(\%) & Length & & T1 & T2 \\
\noalign{\smallskip}\hline\noalign{\smallskip}
\textbf{Scenario 1} & \\[1.ex]
$\quad \mbox{E=1}$
& Global            & 20800.0 & 20796.6 & & 0.036 & 0.068 & 98.16 & 0.612 & & 2673 & 2673 \\
& Disjoint          & 20818.0 & 20801.7 & & 0.043 & 0.077 & 99.24 & 0.751 & &  178 &  406 \\
& 1st order neighb. & 20813.2 & 20798.2 & & 0.043 & 0.073 & 99.21 & 0.743 & &  292 &  546 \\
& 2nd order neighb. & 20812.2 & 20798.1 & & 0.043 & 0.071 & 99.09 & 0.733 & &  413 &  750 \\[2.ex]
$\quad \mbox{E=5}$
& Global            & 35113.7 & 35105.4 & & 0.028 & 0.058 & 98.69 & 0.423 & & 1811 & 1811 \\
& Disjoint          & 35135.5 & 35114.1 & & 0.029 & 0.052 & 98.60 & 0.417 & &  189 &  436 \\
& 1st order neighb. & 35126.4 & 35106.0 & & 0.029 & 0.052 & 98.93 & 0.428 & &  293 &  581 \\
& 2nd order neighb. & 35133.6 & 35114.8 & & 0.029 & 0.054 & 98.82 & 0.441 & &  378 &  724 \\[2.ex]
$\quad \mbox{E=10}$
& Global            & 40846.5 & 40825.7 & & 0.023 & 0.052 & 98.67 & 0.358 & & 1799 & 1799 \\
& Disjoint          & 40864.1 & 40832.0 & & 0.023 & 0.044 & 98.49 & 0.328 & &  182 &  417 \\
& 1st order neighb. & 40849.4 & 40817.0 & & 0.023 & 0.046 & 99.00 & 0.347 & &  277 &  554 \\
& 2nd order neighb. & 40861.6 & 40831.4 & & 0.023 & 0.048 & 98.99 & 0.362 & &  303 &  578 \\[2.ex]
$\quad \mbox{E=50}$
& Global            & 54166.5 & 54050.4 & & 0.014 & 0.039 & 98.29 & 0.239 & & 1866 & 1866 \\
& Disjoint          & 54108.6 & 54003.7 & & 0.013 & 0.032 & 98.33 & 0.205 & &  155 &  348 \\
& 1st order neighb. & 54083.9 & 53970.6 & & 0.013 & 0.034 & 98.81 & 0.219 & &  181 &  371 \\
& 2nd order neighb. & 54109.6 & 53997.3 & & 0.013 & 0.035 & 98.80 & 0.228 & &  244 &  458 \\[2.ex]
\textbf{Scenario 2} & \\[1.ex]
$\quad \mbox{E=1}$
& Global            & 19815.1 & 19810.3 & & 0.048 & 0.109 & 99.80 & 0.811 & & 1609 & 1609 \\
& Disjoint          & 19894.2 & 19874.8 & & 0.070 & 0.127 & 99.51 & 0.904 & &  151 &  340 \\
& 1st order neighb. & 19875.1 & 19856.4 & & 0.062 & 0.120 & 99.78 & 0.907 & &  215 &  410 \\
& 2nd order neighb. & 19868.3 & 19850.4 & & 0.058 & 0.117 & 99.89 & 0.910 & &  284 &  515 \\[2.ex]
$\quad \mbox{E=5}$
& Global            & 34236.2 & 34193.6 & & 0.028 & 0.077 & 99.79 & 0.535 & & 1922 & 1922 \\
& Disjoint          & 34279.1 & 34231.5 & & 0.035 & 0.080 & 99.70 & 0.527 & &  146 &  327 \\
& 1st order neighb. & 34253.1 & 34201.7 & & 0.031 & 0.077 & 99.85 & 0.536 & &  187 &  379 \\
& 2nd order neighb. & 34250.7 & 34197.9 & & 0.030 & 0.077 & 99.87 & 0.541 & &  254 &  476 \\[2.ex]
$\quad \mbox{E=10}$
& Global            & 40028.0 & 39942.7 & & 0.022 & 0.067 & 99.77 & 0.439 & & 1915 & 1915 \\
& Disjoint          & 40055.3 & 39973.9 & & 0.028 & 0.067 & 99.64 & 0.421 & &  136 &  303 \\
& 1st order neighb. & 40025.9 & 39935.9 & & 0.024 & 0.065 & 99.83 & 0.431 & &  166 &  334 \\
& 2nd order neighb. & 40027.8 & 39934.5 & & 0.024 & 0.065 & 99.85 & 0.436 & &  231 &  425 \\[2.ex]
$\quad \mbox{E=50}$
& Global            & 53403.9 & 53086.7 & & 0.013 & 0.047 & 99.55 & 0.269 & & 1885 & 1885 \\
& Disjoint          & 53376.0 & 53105.5 & & 0.015 & 0.044 & 99.53 & 0.253 & &  113 &  247 \\
& 1st order neighb. & 53352.5 & 53054.5 & & 0.013 & 0.044 & 99.64 & 0.260 & &  152 &  302 \\
& 2nd order neighb. & 53366.9 & 53057.9 & & 0.013 & 0.045 & 99.66 & 0.260 & &  219 &  396 \\
\noalign{\smallskip}\hline
\end{tabular}}
\end{table}

\section{Data analysis: colorectal cancer mortality in Spain}
\label{sec:DataAnalysis}
In this section, male colorectal cancer mortality data in the $n=7,907$ municipalities of continental Spain (excluding Baleares and Canary Islands and the autonomous cities of Ceuta and Melilla) are analyzed using the new model proposals. According to recent studies \cite{ferlay2018cancer}, colorectal cancer was the second cause of cancer deaths in male population in Europe (representing 12\% of all cancers deaths) and in Spain in 2018 after lung cancer.
A total of 81,934 colorectal cancer deaths (corresponding to International Classification of Diseases-10 codes C18-C21) were registered for male population in the municipalities of continental Spain during the period 2006-2015, which represents an overall crude rate of 38.54 deaths per 100,000 male inhabitants.
The indirect age-standardization method has been used to compute the number of expected cases using 5-years age groups (internal standardization). This method allows us to compare the relative risk of each municipality with the whole of Spain during the study period. The expected number of cases ranges from 0 to 6,129 (with mean and median values of 1.8 and 10.4, respectively), while the number of observed cases varies from 0 to 5,814 (with mean and median values of 2.0 and 10.4, respectively).

As in the simulation study, the \textit{Global model}, the \textit{Disjoint model}, and $k=1,2,3$ \textit{order neighbourhood models} have been fitted with \texttt{R-INLA} using the $D=15$ Autonomous Regions of Spain as a partition of the spatial domain. The same hyperprior distributions described in Section \ref{sec:SimulationStudy} have been also considered here. Results are shown in \autoref{tab:DataAnalysis_Results}. The computational time for the scalable model proposals are divided into: 1) \textit{running time}, which corresponds to the maximum time of the $D=15$ submodels (that is, assuming that all models have been simultaneously fitted), and 2) \textit{merging time}, corresponding to the computation of the mixture distribution of the risks and the approximate DIC and WAIC values. As expected, the complexity and computational time of the models increases as higher values of neighbourhood order are considered. The largest values of $n_d$ (number of areas for each subdivision) corresponds to the autonomous region of Castilla y León, located at the north-west of Spain, with a total of 2245, 2451, 2744 and 3047 municipalities for neighbourhood models with $k=0 \mbox{ (Disjoint model)},1,2,\mbox{ and } 3$ respectively.

\begin{table}
\caption{Model selection criteria ($\overline{D({\ttheta})}$: mean deviance, $p_{D}$: effective number of parameters, DIC: deviance information criterion, WAIC: Watanaba-Akaike information criterion), computational time (T.run: running time, T.merge: merging time, T.tot: Total time) in seconds and dimension of the data ($n=\sum_{d=1}^D n_d$).}
\label{tab:DataAnalysis_Results}
\resizebox{\textwidth}{!}{
\begin{tabular}{lrrrrrrrr}
\hline\noalign{\smallskip}
Model & $\overline{D({\ttheta})}$ & $p_{D}$ & DIC & WAIC & T.run & T.merge & T.total & $n$ \\
\noalign{\smallskip}\hline\noalign{\smallskip}
Global                  & 26667.6 & 548.5 & 27216.1 & 27237.9 & 1929 & $-$ & 1929 & 7907  \\
Disjoint                & 26510.7 & 656.8 & 27167.5 & 27166.7 &  110 &  26 &  136 & 7907  \\
1st order neighbourhood & 26533.5 & 634.2 & 27167.6 & 27170.5 &  132 &  63 &  195 & 8979  \\
2nd order neighbourhood & 26557.9 & 616.5 & 27174.3 & 27183.3 &  166 &  83 &  249 & 10646 \\
3rd order neighbourhood & 26586.0 & 583.0 & 27169.0 & 27175.4 &  219 & 107 &  326 & 12553 \\
\noalign{\smallskip}\hline
\end{tabular}}
\end{table}

\begin{figure}
\vspace{-0.3cm}
\includegraphics[width=1.1\textwidth]{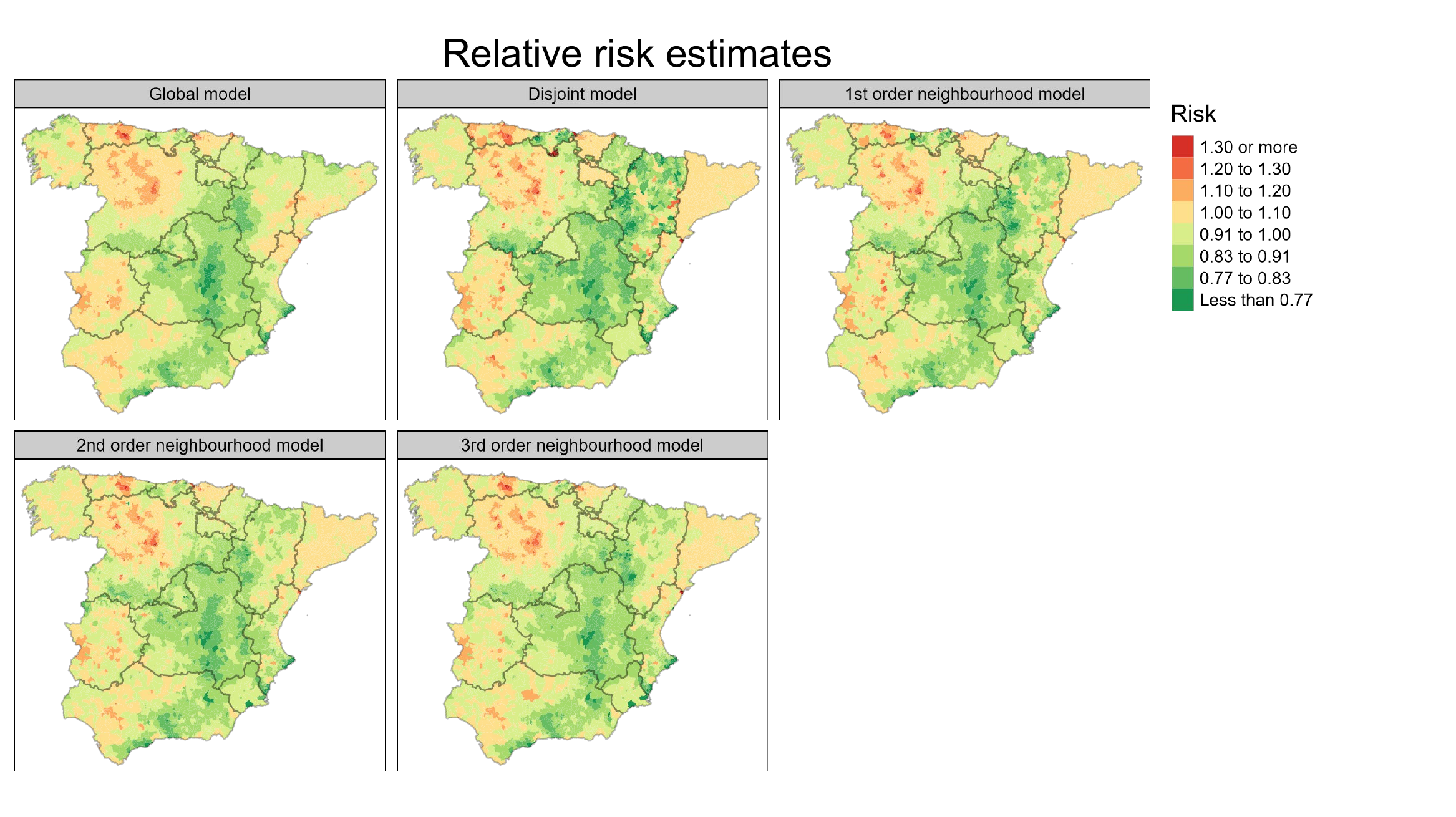}\\
\vspace{-0.3cm}
\caption{Maps of posterior median estimates for $r_i$ of male colorectal cancer mortality data in Spanish municipalities during the period 2006-2015.}
\label{fig:DataAnalysis_Risks}
\includegraphics[width=1.1\textwidth]{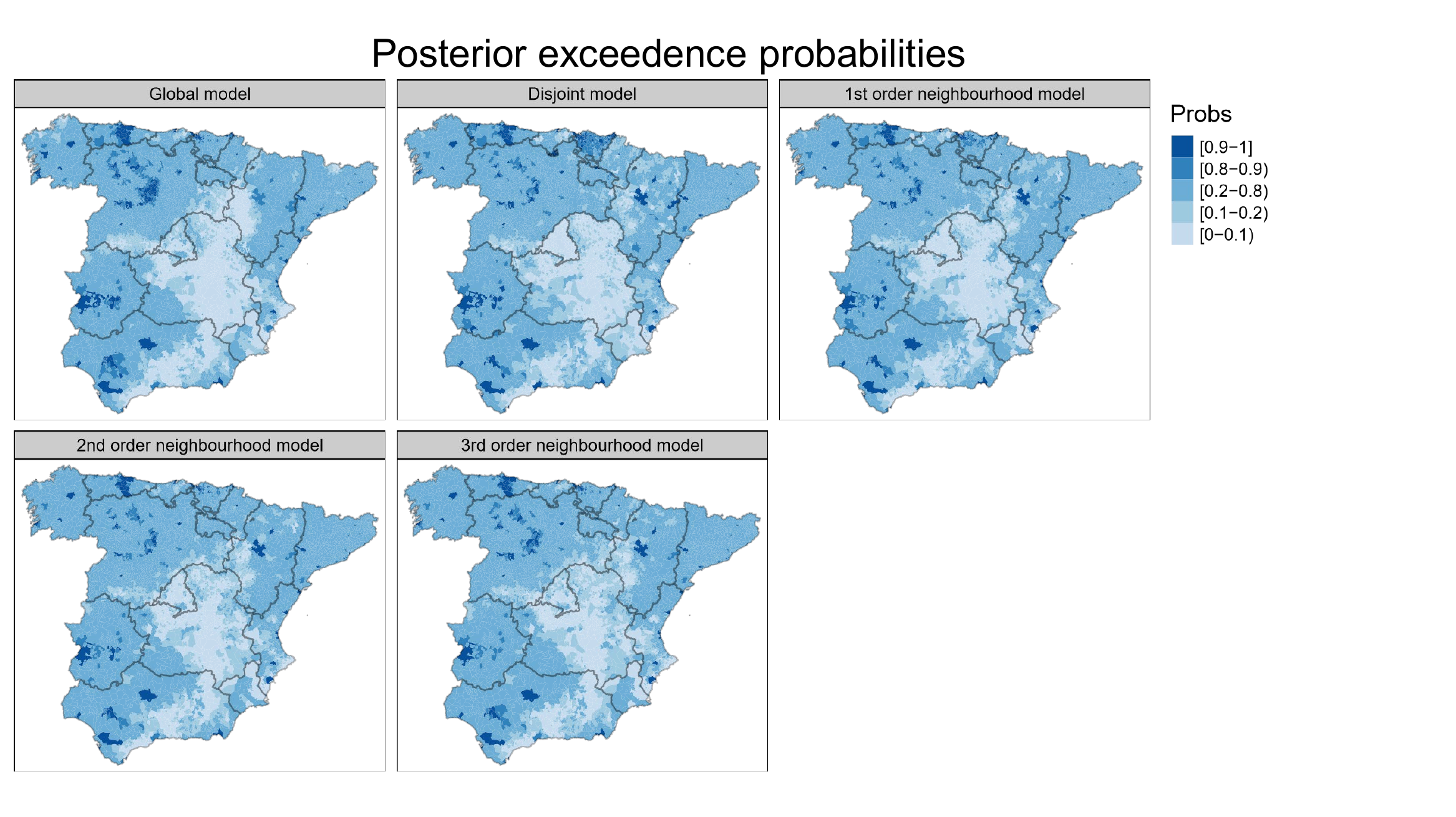}\\
\vspace{-0.3cm}
\caption{Maps of posterior exceedence probabilities $P(r_i>1 | {\bf O})$ of male colorectal cancer mortality data in Spanish municipalities during the period 2006-2015.}
\label{fig:DataAnalysis_Probs}
\end{figure}

Besides the significant reduction in the computational time required to fit the models in INLA, the model selection criteria suggest that the new model proposals outperform the \textit{Global model} in this real data analysis. The maps with posterior median estimates of $r_i$, and posterior exceedence probabilities $P(r_i>1 | {\bf O})$ of male colorectal cancer mortality risks are shown in \autoref{fig:DataAnalysis_Risks} and \autoref{fig:DataAnalysis_Probs}.
In general, very similar spatial patterns are observed for all the models, but \textit{2nd} and \textit{3rd order neighbourhood models} seem to show the most similar risks to those estimated by the \textit{Global model}. Even though small differences are observed in DIC and WAIC values between the scalable model proposals, a greater variability in the degree of spatial smoothness among autonomous regions is observed for the \textit{Disjoint model}, which in some regions as Madrid or Arag\'on leads to not very reasonable relative risk estimates. As expected, this effect seems to be corrected when including neighbouring areas to the spatial subdomains in the \textit{k-order neighbourhood models}.

\section{Discussion and conclusions}
\label{sec:Discussion}

In geostatistics, there are several proposals to deal with massive datas sets. However, the existing methods for analyzing high-dimensional areal count data are still very limited. In this work, we develop a scalable Bayesian model for smoothing mortality or incidence risks in spatial disease mapping when the number of small areas is very large. We propose to divide the main spatial domain into subregions so that local spatial models can be simultaneously fitted reducing the computational time substantially. Although the methodology described in this paper is focused on the INLA estimation strategy, it can  also be adapted to other Bayesian fitting techniques.

As stated, the new proposals require to define a partition of the spatial domain as a first step. A natural choice for this partition are the administrative divisions of the area of interest (such as provinces, states or local health areas). However, if the user has no idea on how to define this initial partition,  a random partition can be also considered by defining a grid over the associated cartography with a certain number of rows and columns (see the vignette accompanying the \texttt{bigDM} package for further details). In a second stage, we propose to fit independent hierarchical Bayesian models including spatially structured and unstructured random effects to smooth the risks in each subregion. Here, two different modeling approaches are defined: a \textit{Disjoint model} where each geographical unit is contained into a single subregion, and a \textit{k-order neighbourhood model} where an overlapping set of regions are defined by adding neighbouring areas to those regions located in the border of the partition. This second approach allows us to eliminate the independence assumption between areas belonging to different subregions, avoiding border effects. Finally, the results of the models are merged to obtain a unique risk estimate for each areal unit. For the \textit{k-order neighbourhood model}, we propose to use a mixture distribution of the estimated posterior probability density functions using the CPO's to compute the mixture weights. In addition, approximations to model selection criteria such as DIC and WAIC are also derived for the scalable models proposed in this paper.

Both the simulation study and the real data analysis indicate that the new methodology provides reliable risk estimates with a substantial reduction in computational time. Moreover, the scalable model proposals avoid the high RAM/CPU memory usage when analyzing massive spatial data. In those cases where small differences in model selection criteria are observed between the \textit{Disjoint} and \textit{k-order neighbourhood model}, we recommend to use the \textit{k-order neighbourhood model} to avoid overfitting and border effects. 

Finally, we think that a great potential of this methodology is its extension to the spatio-temporal setting. The complexity inherent to spatio-temporal interaction models and the even higher dimensionality associated to this type of data, makes necessary the use of scalable techniques for Bayesian inference in small area data. We are currently investigating this issue.


\section*{Acknowledgements}

This research has been supported by the Spanish Ministry of Science and Innovation (project  MTM 2017-82553-R (AEI/FEDER, UE)).  It has also been partially funded by la Caixa Foundation (ID 1000010434), Caja Navarra Foundation, and UNED Pamplona, under agreement LCF/PR/PR15/51100007 (project REF P/13/20).

\cleardoublepage
\bibliographystyle{apalike}
\addcontentsline{toc}{chapter}{References}
\bibliography{biblio}

\end{document}